# STABILITY OF VORTEX PHASES IN FERROELECTRIC EASY-PLANE NANO-CYLINDERS

L. LAHOCHE[1], I. LUK'YANCHUK[2], G. PASCOLI[3]

[1]*Roberval Laboratory, University of Technology of Compiegne, France*

[2]*Condensed Matter Physics Laboratory, University of Picardie, Amiens, France*

[3]*Department of Physics, Physical Faculty, University of Picardie, Amiens, France*

*Bounded charges induced by the polarization gradient in finite-size ferroelectrics are known to produce the unfavorable depolarization electric field that suppresses the uniform ferroelectric state. To reduce the depolarization energy the non-uniform vortex (toroidal) state is formed inside ferroelectric nano-particles, nano-disks, and nano-rods. Based on self-consistent solution of Ginzburg-Landau equations coupled with electrostatic equations, we study the multivortex toroidal states appearing in the nanometric easy-plane ferroelectric cylinder. The geometrical textures, critical temperatures and stability regions for these states are calculated.*

## I. Introduction

During last decades the numerous studies of the influence of finite size effects on the properties of ferroelectrics such as Curie temperature, hysteresis of polarization, coercitive field etc. were done [1-3]. In particularly, the drastic decrease of transition temperature in small samples was attributed to the effect of depolarization field that leads to the non-uniform distribution of polarization [4-6]. However, the determination of the critical sample size, below which the ferroelectric states is completely suppressed, is still an open problem. Current experimental techniques have allowed the detection of ferroelectricity in perovskite films down to a thickness of 4 nm [7]. Several theoretical studies, based mostly on first-principles effective Hamiltonian calculations do confirm the existence the ferroelectric order parameter down to the very small sample size : Geneste [1] has shown the presence of ferroelectricity-induced distortion in nano-wires of $BaTiO_3$ below a critical diameter of about 1.2 nm, Junquera [3] has estimated the critical width of ferroelectric films as 2.4 nm whereas Naumov [8] has determined the critical diameter of nanorods and nanodisks to be close to 3.2nm.

The objective of the present work is to investigate the properties of ferroelectric nano-cylinders with easy-plane distribution of polarization using both analytical calculations and numerical solution of Ginzburg-Landau equations, coupled with Maxwell electrostatic equations. The great advantage of such phenomenological approach in comparison with first-principles calculations [8] is that, it allows increasing the size of the system (number of elementary cells) by more then five order of magnitude. We shall study the ferroelectric nano-cylinders of diameters from 2 to 50nm, calculate the reduction of the critical temperature and demonstrate the existence of the unconventional multi-vortex toroidal(meta-) stable polarization textures. Note that we consider the simplest 2D situation that corresponds to the strongly anisotropic uniaxial easy-plane displacive ferroelectrics (like e.g. Tetragonal Tungsten Bronze compounds). The cubic Perovskite-type ferroelectrics can have more complex vortices with polarization escape in the third dimension [9]. Such compounds are not considered here.

## 2. Description of the Model

Cross section of easy-plane isotropic ferroelectric cylinder of radius R, embedded into paraelectric media is shown in Fig.1. The spatial distribution of polarization P and electric potential $\varphi^{(f)}$ in the ferroelectric region ($\Omega_f$) are described by the coupled electrostatic and nonlinear Ginsburg-Landau equations [10,11]

$$\nabla^2 \varphi^{(f)} = 4\pi \nabla P, \qquad (1)$$

$$4\pi \varepsilon_{//}^{-1} [(t + P^2)P - \xi_0^2 \nabla^2 P] = -\nabla \varphi^{(f)}, \qquad (2)$$

where the reduced temperature t is expressed via the bulk critical temperature $T_{c0}$ as $t=T/T_{c0}-1$, polarization P is measured in units of uniform polarization $P_0$ at $t=-1$, the dimensionless parameter $\varepsilon_{//} \gg 1$ is expressed via the Curie constant: $\varepsilon_{//} = C/T_{c0}$, $E = -\nabla \varphi$ and $\nabla^2 = \partial_x^2 + \partial_y^2$. For simplicity we assume the in-plane isotropy of the system.

In the paraelectric region ($\Omega_p$), the electric potential $\varphi^{(p)}$ satisfies the Laplace equation:

$$\nabla^2 \varphi^{(p)} = 0. \qquad (3)$$

The electrostatic boundary conditions at the para/ferro interface ($\partial \Omega_1$) are given by:

$$(n.\nabla)(\varphi^{(f)} - \varphi^{(p)}) = 4\pi n.P, \qquad (4)$$

$$(n.\nabla)P = 0, \qquad (5)$$

$$\varphi^{(f)} = \varphi^{(p)}, \qquad (6)$$

where the unit vector n is normal to the boundary $\partial \Omega_1$ and directed toward ferroelectric region. At the boundary $\partial \Omega_2$, the electrostatic potential is supposed to be zero:

$$\varphi^{(p)} = 0. \qquad (7)$$

## 3. Transition temperature: Linear problem

Since ferroelectric polarization vanishes close to transition temperature $t_c$ the nonlinear term $P^3$ in equation (2) can be neglected. The corresponding linearized equation

$$4\pi\varepsilon_{//}^{-1}[(tP - \xi_0^2\nabla^2 P] = -\nabla\varphi^{(f)}, \qquad (8)$$

defines the transition temperature and the polarization texture close to $t_c$. To calculate $t_c$ we decompose the polarization onto gradient and rotational terms as

$$\overline{P} = \overline{P}_t + \overline{P}_e = \text{rot}\Psi e_z + \nabla\Phi = \nabla\Psi \times e_z + \nabla\Phi \qquad (9)$$

(here $4\pi P \to \overline{P}$).

Now the equation (8) can be decomposed as:

$$(t - \nabla^2)\nabla\Psi = 0, \qquad (10)$$

$$(t - \nabla^2)\nabla\Phi = -\varepsilon_{//}\nabla\varphi^{(f)}, \qquad (11)$$

And the equation (1) is written as :

$$\nabla^2\varphi^{(f)} = \nabla^2\Phi, \qquad (12)$$

Solution of equation (12) can be written in the form $\varphi^{(f)} = \Phi + \Sigma$, that permits to present the system (10),(11) as :

$$(\nabla^2 + K_2^2)(\Phi - \varepsilon_{//}\frac{\Sigma}{K_2^2}) = 0, \qquad (13)$$

$$(\nabla^2 + K_1^2)\Psi = 0. \qquad (14)$$

where $K_1^2 = -t$, $K_2^2 = -t + \varepsilon_{//}$.

In cylindrical coordinates ($\rho,\theta$) the general solution of this system can be written as:

$$\Psi = \Sigma(A_m\cos m\theta + B_m\sin m\theta)J_m(k_1\rho), \qquad (15)$$

$$\Phi = \Sigma[C_m\frac{J_m(k_2\rho)}{J_m(k_2R)} + \varepsilon_{//}\frac{E_m}{K_2^2}(\frac{\rho}{R})^m]\cos m\theta + [D_m\frac{J_m(k_2\rho)}{J_m(k_2R)} + \varepsilon_{//}\frac{E_m}{K_2^2}(\frac{\rho}{R})^m]\sin m\theta, \qquad (16)$$

$$\varphi^{(f)} = \Sigma[C_m\frac{J_m(k_2\rho)}{J_m(k_2R)} + \varepsilon_{//}E_m\frac{K_1^2}{K_2^2}(\frac{\rho}{R})^m]\cos m\theta + [D_m\frac{J_m(k_2\rho)}{J_m(k_2R)} + \varepsilon_{//}F_m\frac{K_1^2}{K_2^2}(\frac{\rho}{R})^m]\sin m\theta, \qquad (17)$$

$$\varphi^{(p)} = \Sigma(G_m \cos m\theta + H_m \sin m\theta)(\frac{R}{\rho})^m. \tag{18}$$

where $J_m(x)$ are the cylindrical Bessel functions, and $A_m$, $B_m$, $C_m$, $D_m$ are the arbitrary coefficients defined by the boundary conditions. The compatibility condition for the system (15)-(18) leads to the characteristic equation:

$$\Lambda = \begin{vmatrix} -J_m(x) & 1 & \varepsilon_{//} + 2t \\ m(xJ'_m(x) - J_m(x)) & y^2 \frac{I'_m(y)}{I_m(y)} & -m(m-1)\varepsilon_{//} \\ x^2 J''_m(x) & m(y\frac{I'_m(y)}{I_m(y)} - 1) & -m(m-1)\varepsilon_{//} \end{vmatrix} = 0. \tag{19}$$

that defines the critical temperature of the system. Here: $x = \sqrt{-t}.R$, $y = \sqrt{\varepsilon_{//} - t}.R$ and $I_m(x)$ are the hyperbolic Bessel functions.

Numerical solution of equation (19) was done for different m to identify the critical temperature $t_c$ vs. radius R. The material constants were selected as: $\varepsilon_{//}=C/T_{c0}=500$, $\xi_0=10A$ that corresponds to the realistic situation of displacive ferroelectrics. It was shown that for each m there exists a series of solutions of (19), classified by the integer numbers p=1,2,…, having different critical temperatures. Dependencies of critical temperatures on R for different m for p=1 and p=2 are shown in Fig. 2. For instance, for the cylinder of radius R=10nm, the modes with p=1 and m=0, m=1, m=2, m=3, have the critical temperatures t=-0.03, t=-0.13, t=-0.24 et t=-0.39. The temperature of ferroelectric transition $t_c(R)$ is given by the maximal critical temperature and corresponds to the mode m=0, p=1. Note that this mode remains stable even for the very small values of R ($R_c$=1.91nm at t=-1).

**4. Nonlinear problem: Results of numerical calculations**

Complete numerical analysis of the 2D nonlinear problem (equations (1)- (7)) was performed using the Comsol Multiphysics [12] finite-element toolbox that permitted to follow the polarization and electric field space distribution as a function of radius 0<R<50nm and temperature t.

Fig. 3 shows the distribution of polarizations for different modes (p,m) that do correspond to the multivortex toroidal states. Although the mode p=1, m=0 is the most stable one, other modes corresponds to the local energy minima and can be observed as the metastable state as a result of the rapid quench. The region of the metastability for each mode is presented in Fig. 2. Note that at lower temperature the sharp domain walls that separate the cylinder onto different ferroelectric domains appear. The detailed analysis of their temperature evolution will be published elsewhere.

This work was supported by region of Picardie, France and by European F6 project "Multiceral".

**Figures**

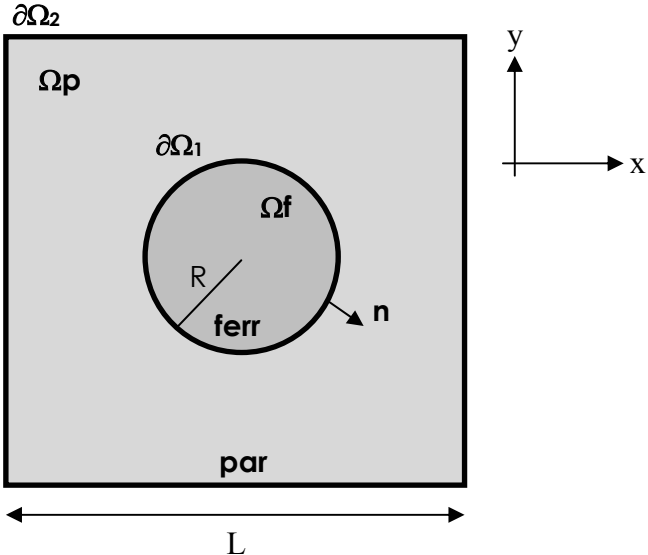

Fig.1. : Geometry of the problem : cross section of easy-plane ferroelectric cylinder embedded into paraelectric material.

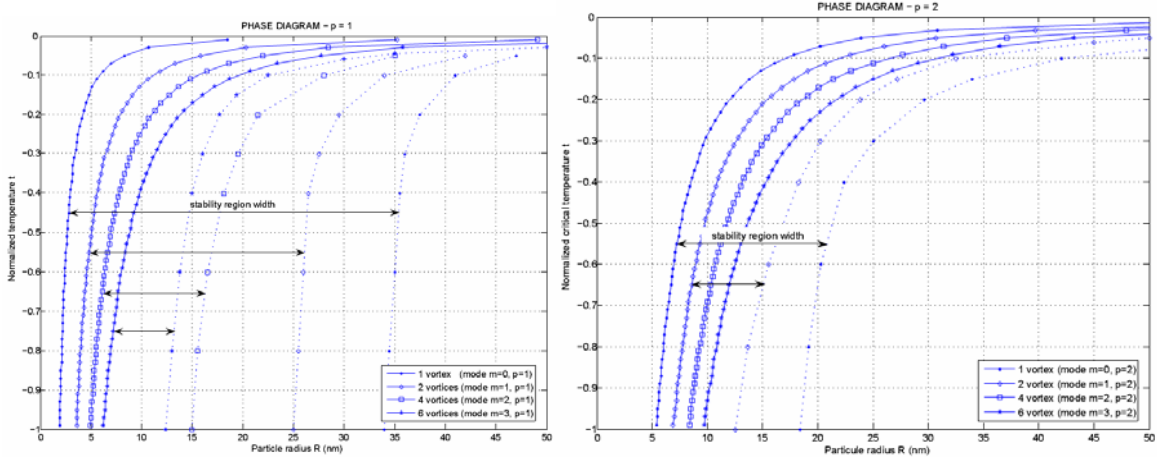

Fig.2. : Phase diagram for the modes p=1 (left) and p=2 (right)

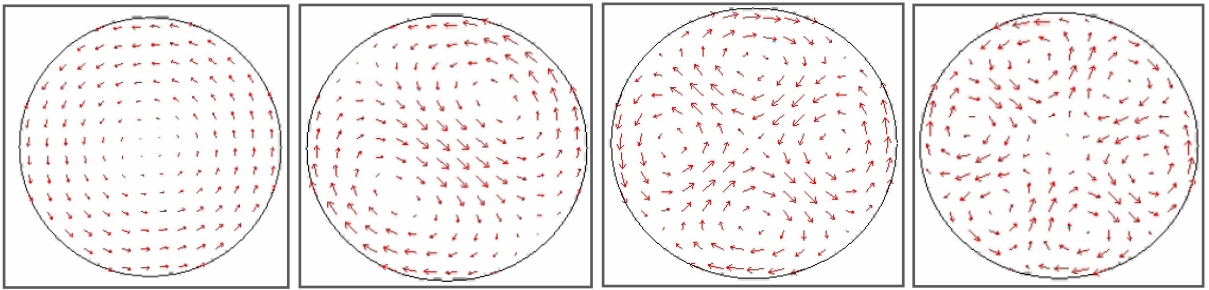

Fig.3a : Modes p=1, m=0,1,2,3

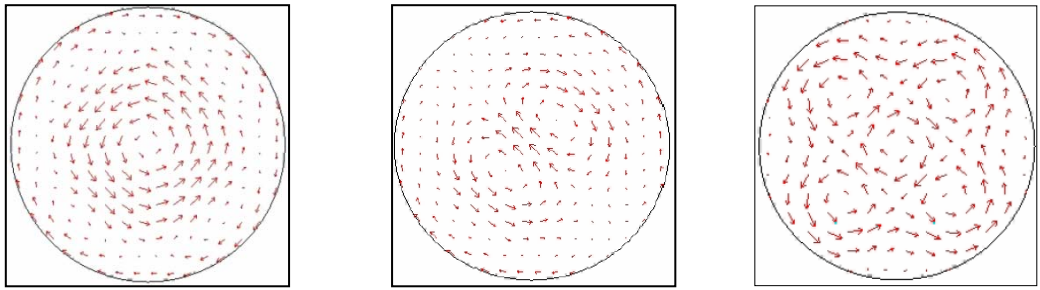

Fig.3b : Modes p=2; m=0,1,2